\documentclass[a4paper,fleqn,usenatbib]{mnras}

\usepackage[T1]{fontenc}
\usepackage{ae,aecompl}

\usepackage{graphicx}	
\usepackage{amsmath}	
\usepackage{amssymb}	
\usepackage{hyperref}

\title[The resolved SFH of M51a]{The resolved star formation history of M51a through successive Bayesian marginalization}

\author[Mart\'{\i}nez-Garc\'ia et al.]{
Eric E. Mart\'inez-Garc\'ia,$^{1}$\thanks{E-mail: martinezgarciaeric@gmail.com}
Gustavo Bruzual,$^{2}$
Gladis Magris C.,$^{3}$
\newauthor
and Rosa A. Gonz\'alez-L\'opezlira,$^{2,4,5}$
\\
$^{1}$Cerrada del Rey 40-A, Chimalcoyoc Tlalpan, Ciudad de M\'exico, C.P. 14630\\
$^{2}$Instituto de Radioastronom\'ia y Astrof\'isica, UNAM, Campus Morelia,
      Michoac\'an, M\'exico, C.P. 58089\\
$^{3}$Centro de Investigaciones de Astronom\'ia, Apartado Postal 264, M\'erida 5101-A, Venezuela\\
$^{4}$Argelander Institut f\"ur Astronomie, Universit\"at Bonn, Auf dem H\"ugel 71, D-53121 Bonn, Germany\\
$^{5}$Helmholtz-Institut f\"ur Strahlen-und Kernphysik (HISKP), Universit\"at Bonn, Nussallee 14-16, D-53115 Bonn, Germany
}

\date{Accepted XXX. Received YYY; in original form ZZZ}

\pubyear{2018}

\begin{document}
\label{firstpage}
\pagerange{\pageref{firstpage}--\pageref{lastpage}}
\maketitle

\begin{abstract}

We have obtained the time and space-resolved star formation history (SFH)
of M51a (NGC~5194) by fitting {\it GALEX}, SDSS, and near infrared pixel-by-pixel
photometry to a comprehensive library of stellar population synthesis models drawn
from the Synthetic Spectral Atlas of Galaxies (SSAG).
We fit for each space-resolved element (pixel) an independent model where
the SFH is averaged in 137 age bins, each one 100 Myr wide.
We used the Bayesian Successive Priors (BSP) algorithm
to mitigate the bias in the present-day spatial mass distribution.
We test BSP with different prior probability distribution functions (PDFs);
this exercise suggests that the best prior PDF is the one concordant
with the spatial distribution of the stellar mass as inferred
from the near infrared images.
We also demonstrate that varying the implicit prior PDF
of the SFH in SSAG does not affects the results.
By summing the contributions to the global star formation rate of each pixel,
at each age bin, we have assembled the resolved star formation history
of the whole galaxy. According to these results, the star formation rate
of M51a was exponentially increasing for the first 10 Gyr after the Big Bang,
and then turned into an exponentially decreasing function until the present day.
Superimposed, we find a main burst of star formation at $t\approx$ 11.9 Gyr
after the Big Bang.

\end{abstract}




\section{Introduction}

Modern multiwavelength and integral field unit (IFU)
data provide unprecedented information that can help uncover essential
aspects of galaxy formation and evolution.
Resolved studies of galaxy properties on a pixel-by-pixel~\citep[e.g.][]{zcr09,men12,sor15,dia15,mar17,abd17},
or spaxel-by-spaxel basis~\citep[e.g.][]{gond14,gond15,gond16,can16,iba16,deAmo17}
have become standard practice in modern investigations.
These methods commonly use state-of-the-art stellar population synthesis (SPS)
models to compare with observations.
\citet{mar17} introduced a novel and sophisticated Bayesian fitting method
to remove a bias originated from traditional fitting techniques to pixel-by-pixel
photometry. The bias consists in an apparent spatial coincidence between the resolved
stellar mass surface density (or mass-map) and the observed dust lanes
(produced by internal galaxy extinction), resulting in an illusory filamentary
spatial distribution for the mass.
In this work we aim to describe an ancillary output of the~\citet{mar17} mass-map method,
related to the recovery of the star formation history (SFH) of a disk galaxy.
The novel method determines the resolved SFH by fitting a library of SPS models to UV,
optical, and near infrared (NIR) photometry on a pixel-by-pixel basis.
The recent star formation rate (SFR, $\Psi$) of a galaxy can be inferred from its ultraviolet
(UV) continuum luminosity. However, $\Psi_{\rm UV}$ may exceed other $\Psi$
indicators if the extinction is overestimated due to the
underlying Balmer absorption~\citep{ros02}. 
\citet{sal07,sal16} have shown that $\Psi$ can be better estimated from spectral
energy distribution (SED) fits that include the optical and NIR
ranges besides the UV. We have included these wavelength ranges in our method
and applied it to the Whirlpool galaxy, aka M51a.\footnote{The RC3 type of M51a is SA(s)bc~pec~\citep{deV91}.}
The results and their implications are shown in this paper.

Throughout this work we adopt the cosmological parameters of~\citet{ben14},
and use the cosmological calculator of~\citet{wri06} where necessary.

\section{Observational data}

For this investigation we use far and near ultraviolet
($FUV$ and $NUV$, respectively) imaging from the {\it GALEX}
Ultraviolet Atlas of Nearby Galaxies~\citep{gil07},
$g$- and $i$-band optical data from the 12th Sloan Digital Sky
Survey (SDSS) data release~\citep{ala15},
and the NIR $K_{s}$ mosaic from~\citet{gon96}.
The latter was photometrically calibrated using the
Two Micron All Sky Survey~\citep[2MASS,][]{skr06}.

We registered the images to the spatial resolution
of the $FUV$ image (which has a plate scale of $1.5\arcsec$ pixel$^{-1}$)
using the world coordinate system (WCS) information.
We noticed a spatial mismatch between the structural features and star forming
regions of the $NUV$ and the $g$ image that
is not present when the $FUV$ and $g$ images are compared.
We correct for this discrepancy by registering the $NUV$ to the $FUV$ image
based on the features they have in common.
Foreground stars were removed by replacing their pixels with
values similar to the background-subtracted `sky'.

The $FUV$ and $NUV$ images have a point-spread function (PSF) with a
full width at half maximum (FWHM) a factor of $\sim$4 larger than SDSS frames.
In order to get a common PSF for all our data,
we use the 2017 version of the convolution kernels of~\citet{ani11}.

To increase the signal to noise (S/N) ratio of the outer
disk pixels in the $NUV$ frame, we adopt the~{\sc{Adaptsmooth}} code of~\citet{zbt09}.
We require a minimum S/N ratio per pixel of 20,
a maximum smoothing radius of 10 pixels, and assume background-dominated noise.
The code produces a `smoothing mask', which contains the smoothing
radius (in pixels) at each position. We then use this output mask
in a second run of~{\sc{Adaptsmooth}} on the $FUV$, $g$, $i$,
and $K_{s}$ data. Subsequently, we compute the 1-$\sigma_{\rm mag}$ error,
on a pixel-by-pixel basis,
for each band, following~\citet[][their section 4]{mar17}.
We assume zero-point errors of
$\sigma_{\rm calib}\approx0.15$ mag for the $FUV$ and $NUV$ data,
$\sigma_{\rm calib}\approx0.01$ mag for the SDSS images, and
$\sigma_{\rm calib}\approx0.033$ mag for the $K_{s}$ mosaic.

We end up with $\sim8\times10^4$ data pixels for each image that can be
fitted with our method.
We assume a distance to M51a of $8.58\pm0.10$ Mpc~\citep{mcq16},
which results in a physical scale of $\sim62$ pc pixel$^{-1}$.
Taking into account the PSF of the $NUV$ band (FWHM$\sim5.25\arcsec$),
the recovered physical scale is $\sim0.22$ kpc.
Optical and NIR Galactic extinction was taken into account
by using the~\citet{sch11} recalibration of~\citet{sch98} with $R_{V}=3.1$.
$NUV$ and $FUV$ Galatic extinction was corrected as
in~\citet[][their equations 4 and 7, respectively]{pee03}.

\section{Analysis}~\label{analysis}

\begin{figure} 
\centering
\includegraphics[width=\columnwidth]{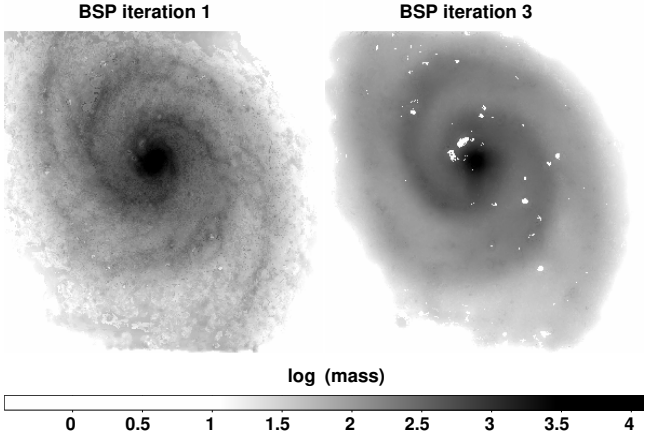}
\caption[fig1.ps]{
{\it{Left}}: M51a stellar mass-map derived with the BSP algorithm, 
iteration number 1,
based on $(FUV-NUV)$, $(NUV-g)$, $(g-i)$ and $(i-K_{s})$ colours,
and $K_{s}$ mass-to-light ratio, $\Upsilon_{*}^{K_{s}}$.
{\it{Right}}: same as left for BSP iteration number 3.
Mass in M$_{\sun}$ pc$^{-2}$.
}~\label{fig1}
\end{figure}

\begin{figure} 
\centering
\includegraphics[width=\columnwidth]{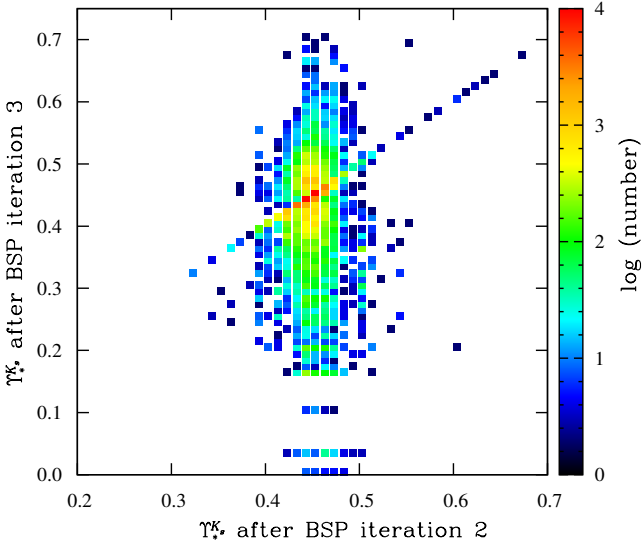}
\caption[fig2.eps]{
2D histogram of the posterior $\Upsilon_{*}^{K_{s}}$ of M51's pixels,
after BSP iterations number 2 and 3, in the $x$- and $y$-axis, respectively.
}~\label{fig2}
\end{figure}

From the observational data we construct the
$(FUV-NUV)$, $(NUV-g)$, $(g-i)$, and $(i-K_{s})$ colour images.
These are used as input to the Bayesian Successive Priors (BSP)
method of~\citet{mar17}. Briefly, the BSP algorithm consists
of three iterations to produce a stellar surface density mass-map consistent
with the NIR spatial structure of a disk galaxy~\citep[see, e.g.,][]{rix93}.
In the first iteration, a {\it maximum-likelihood} estimate is used to
maximise the probability $P\propto\exp \left(-\frac{\chi^{2}}{2}\right)$,
where $\chi$ represents the difference between the observed
colours and the predicted colours of the SPS library,
weighted by their errors. The second iteration assumes a constant
stellar mass-to-light ratio $\Upsilon_{*}$, in the NIR, for the entire disk.
The third iteration deals with the resolved elements (pixels) that
cannot be adequately fitted with a constant $\Upsilon_{*}^{\rm NIR}$.
In this manner, the second and third
iterations introduce a prior probability distribution function (PDF)
to account for the observed NIR spatial structure;
the best fit model is the one that maximises the probability

\begin{equation}~\label{eq_PUpsilon}
P(\Upsilon_{*} \mid C) \propto \exp \left(-\frac{\chi^{2}}{2}\right)
\exp \left(-\frac{1}{2}\left[\frac{\Upsilon_{*}^{\rm prior} - \Upsilon_{*}} 
{\sigma_{\Upsilon_{*}}} \right]^{2}\right),
\end{equation}

\noindent where $C$ represents the observed  colours for a certain stellar population,
and $\sigma_{\Upsilon_{*}}\approx\sigma_{\rm mag} \Upsilon_{*}^{\rm prior}$.
We obtain the resolved stellar mass-map of M51a from the
$(FUV-NUV)$, $(NUV-g)$, $(g-i)$, and $(i-K_{s})$ colours, 
hence $N_{\rm colours}=4$, and
the mass-to-light ratio $\Upsilon^{K_{s}}_{*}$ in the $K_s$ band.
We find that, as seen in Figure~\ref{fig1}, the inclusion of the UV bands in the first iteration fits does
not remove the bias and the spatial structure described by~\citet{mar17}.
Thus, the application of the full BSP algorithm is justified in this work.
We should also mention that although BSP iteration number 2 assumes a prior
$\Upsilon^{K_{s}}_{*}$ that is a constant for the entire disk, the posterior $\Upsilon^{K_{s}}_{*}$
is not necessarily univariate (for both BSP iterations number 2 and 3).
This is shown in Figure~\ref{fig2}, where we plot a 2-D histogram of
the posterior $\Upsilon^{K_{s}}_{*}$ after BSP iterations number
2 and 3, for all the fitted pixels of M51a~\citep[see also][their figure 11, panel d]{mar17}.

\begin{figure} 
\centering
\includegraphics[width=\columnwidth]{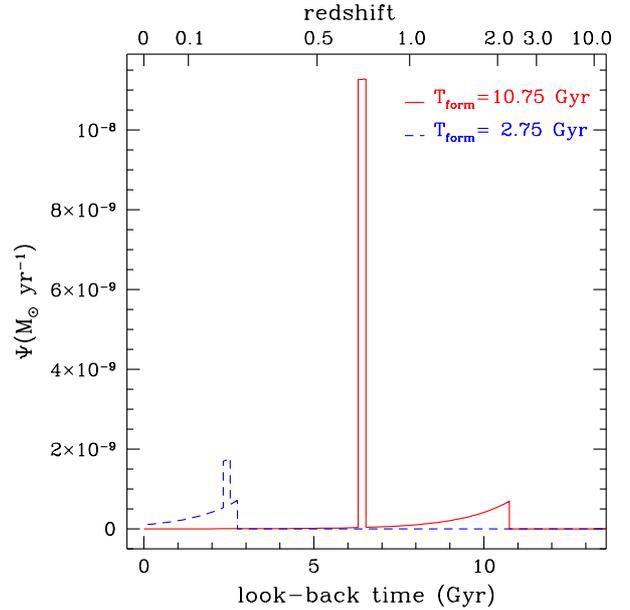}
\caption[fig3.eps]{
SFHs for two different SSAG templates.
{\it Solid red line:}
the galaxy starts forming stars at T$_{\rm form}$=10.8 Gyr ago,
and experiences a burst after 4.2 Gyr, lasting 0.2 Gyr.
{\it Dashed blue line:}
T$_{\rm form}$ = 2.8 Gyr, burst starts after 0.2 Gyr, lasting 0.2 Gyr.
Star formation rates, $\Psi$, scale with galaxy mass~\citep{bru03}.}
~\label{fig3}
\end{figure}

For this research we adopt an SPS library consisting of
$\sim7\times10^4$ templates extracted from the 2017 version of the Synthetic Spectral Atlas of
Galaxies (SSAG-2017 hereafter).\footnote{\url{http://www.astro.ljmu.ac.uk/~asticabr/SSAG.html}}
In the SSAG-2017, the stellar metallicity is distributed uniformly between
$0.006 \leqq Z/Z_{\sun} \leqq 3.53$.
The effects of dust are computed with the model of~\citet{cha00};
the dust parameters follow Gaussian PDFs.
The SSAG spectra are convolved with a Gaussian filter to
mimic the effects of the stellar velocity dispersion.
The adopted stellar initial mass function (IMF) is~\citet{cha03}.
The SSAG uses an SFH recipe proposed by~\citet{che12}.
This SFH prescription consists of a first episode of star formation (SF)
characterised by an exponentially declining event.
For all templates the beginning of star formation is determined
by the parameter T$_{\rm form}$, in look-back time units.
Also, in the SSAG-2017, 55 per cent of the galaxies
experience a superimposed burst of SF of finite duration
and random amplitude~\citep[cf.][their Appendix B]{mag15}.
Bursts are onset at any time in the past,
but are constrained such that 15 per cent of the
total galaxies present a burst in the last 2 Gyr.
Some galaxies may also undergo a `truncation' event, at which $\Psi$
starts to decline at a faster rate than before.
It is assumed that 30 per cent of the galaxies experience this
truncation event, and that 35 per cent of the truncation events occur over the last
2 Gyr (this corresponds to 10 per cent of the total galaxies).
As an example, in Figure~\ref{fig3} we show the SFHs for two SSAG templates.

The BSP algorithm produces a mass-map by 
allocating a probability to each template in the corresponding SPS library.
Each template has its own set of parameters, e.g., metallicity, or dust content.
In this sense, we obtain a resolved map not only for the stellar mass,
but also for each parameter in the library, including the SFH.

For the present analysis, the SFH for each template is averaged in age bins
of 100 Myr, for a total of 137 age bins between today 
and $\sim0.02$ Gyr after the Big Bang.
The average SFR for each bin is then
\begin{equation}
    \langle\Psi(t_{L})\rangle =
     \frac {\int_{t_{L}}^{t_{L}+t_{\rm bw}} \Psi(t') {\rm d}t'}{t_{\rm bw}},
\end{equation}

\noindent where time $t_{L}$ is the look-back time of the bin's boundary
(the side closest to $t_{L}=0$, i.e., today),
and $t_{\rm bw}=10^8$ yr is the corresponding bin width.
We calculate $\langle\Psi(t_{L})\rangle$ for each pixel and then sum
over all pixels to obtain the global value
\begin{equation}~\label{eq_psi_global}
\langle\Psi(t_{L})\rangle^{\rm global} = \sum\limits_{j}\sum\limits_{i} \langle\Psi(t_{L})\rangle_{ij},
\end{equation}

\noindent where $\langle\Psi(t_{L})\rangle_{ij}$ is the average SFR
of the $i^{\rm th},j^{\rm th}$ pixel, for time $t_{L}$.

The uncertainties for $\langle\Psi(t_{L})\rangle^{\rm global}$ are estimated as follows.
For any given parameter $\mathbf{X}$, the error
can be approximated by estimating the 16th and 84th percentiles,
P$_{16}$ and P$_{84}$, respectively, of the resulting probability
distribution for such parameter, and then using

\begin{equation}~\label{sigma_P}
 \sigma_{\mathbf{X}} = ( {\rm P}_{84} - {\rm P}_{16} ) / 2.
\end{equation}

Even when $\langle\Psi(t_{L})\rangle$ for each age bin is not an independent parameter,
we take advantage of the above procedure to approximate the error for 
$\langle\Psi(t_{L})\rangle$ at each given $t_{L}$.
Also, the fitted $\langle\Psi(t_{L})\rangle$ values from the SPS library
need to be scaled by the recovered stellar mass (the one obtained from the fitted mass-to-light
ratio, $\Upsilon_{*}$,
and scaled by the apparent luminosity), since they are obtained with a mass normalisation~\citep{bru03}.
The recovered $\langle\Psi(t_{L})\rangle$
is thus obtained as the product of the fitted $\langle\Psi(t_{L})\rangle$ and the
ratio of the recovered stellar mass to the normalised mass given by the SPS library.
We propagate the error of this product ignoring the correlation terms.
Subsequently, we propagate the errors in the sum given by equation~\ref{eq_psi_global}.

\section{Results}

\begin{figure} 
\centering
\includegraphics[width=\columnwidth]{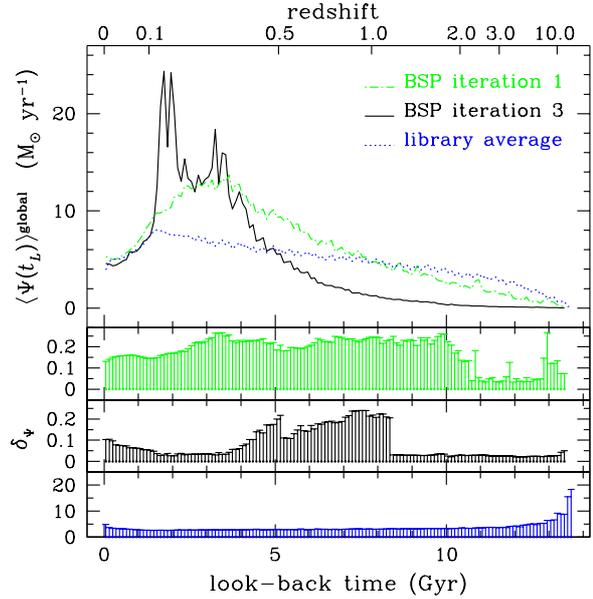}
\caption[fig4.eps]{
Top panel: resolved SFH of M51a, obtained with the BSP algorithm
in age bins 100 Myr wide.
{\it Dashed-dotted green line:} result obtained after BSP iteration number 1.
{\it Solid black line:} result after BSP iteration number 3.
{\it Dotted blue line:} the average of all SFHs in the SSAG-2017 library,
multiplied by the present-day resolved stellar mass.
Second from top panel: relative error, $\delta_{\Psi}=\frac{\sigma_{\Psi}}{\psi}$,
for BSP iteration number 1.
Third from top panel: relative error for BSP iteration number 3.
Bottom panel: relative error for the average of SFHs in the SSAG-2017 library.}
~\label{fig4}
\end{figure}

Figure~\ref{fig4} shows the SFH of M51a resulting after BSP iterations number 1 and 3,\footnote{
Taking into account the error in the distance to the galaxy (not included in the uncertainty)
will change $\Psi$ by $\pm2$ per cent.} with a dashed-dotted (green) and solid (black) lines, respectively.
In this plot the relative error is given by $\delta_{\Psi}=\frac{\sigma_{\Psi}}{\psi}$.
The uncertainty varies with look-back time and is relatively small if the templates that
best fit the data do not show much SF activity at those periods.
Also, $\delta_{\Psi}$ is better constrained when all of the SSAG-2017 templates are used,
as compared to the case when a subset of the SSAG-2017 is adopted.
The two SFHs are quite different, and will be discussed in section~\ref{SFH_paramreti}.
Additionally, in Figure~\ref{fig4} (dotted blue line) we show the average of all SFHs in
our SSAG-2017 library,

\begin{equation}~\label{eq_psi_average}
\langle\Psi(t_{L})\rangle^{\rm library~average} =
\frac{\sum_{j=1}^{N_{\rm templates}} \langle\Psi(t_{L})\rangle_{j}}
{N_{\rm templates}},
\end{equation}

\noindent where $N_{\rm templates}$ stands for the number of templates in our SPS library.
In this case the relative uncertainty represents the ratio
of $\langle\Psi(t_{L})\rangle^{\rm library~average}$ to the standard deviation of the
$\langle\Psi(t_{L})\rangle$ distribution.
The form of the curve for $\langle\Psi(t_{L})\rangle^{\rm library~average}$ is not
correlated to the results obtained with BSP as explained in appendix~\ref{SFH_prior}.

\subsection{Radial gradients in SFHs}

\begin{figure} 
\centering
\includegraphics[width=\columnwidth]{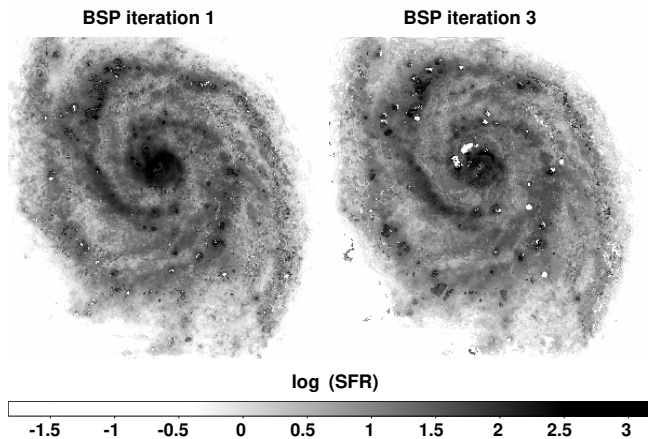}
\caption[fig5.ps]{
{\it{Left}}: M51a SFR-map at $t_{L}=0$ derived with the BSP algorithm, iteration number 1.
{\it{Right}}: same as left panel, for BSP iteration number 3.
SFR in M$_{\sun}$ Gyr$^{-1}$ pc$^{-2}$.
}
~\label{fig5}
\end{figure}

\begin{figure*} 
\centering
\includegraphics[width=1.0\hsize]{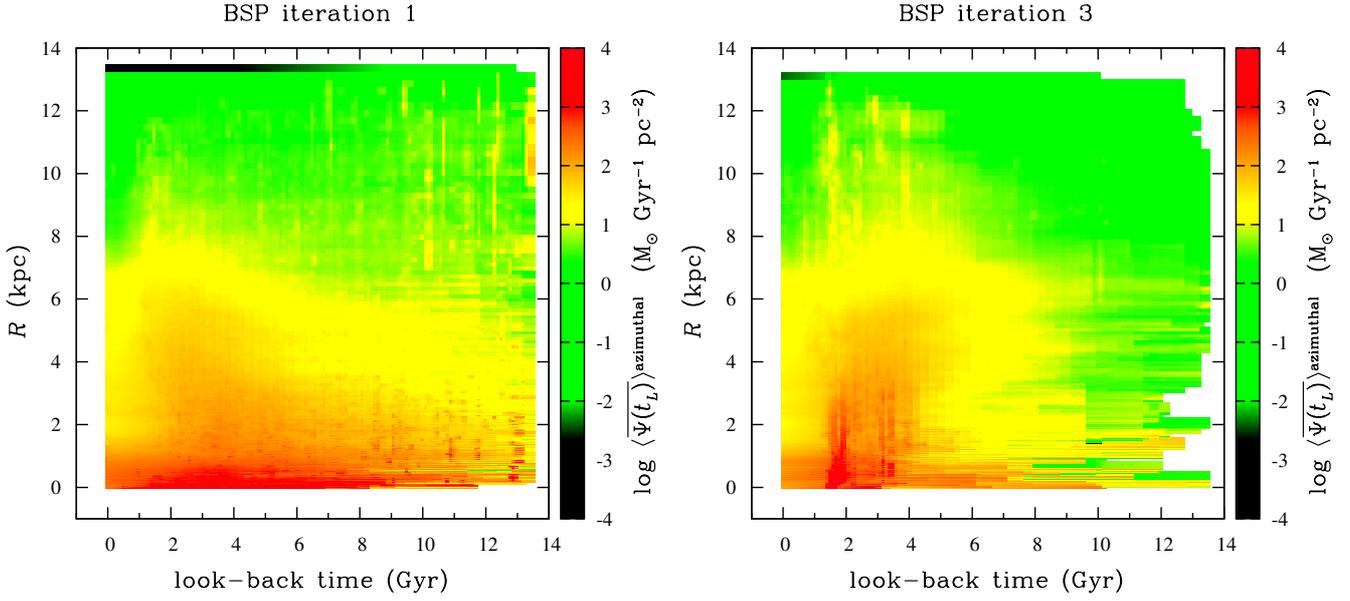}
\caption[fig6.eps]{
Azimuthally averaged SFR $\langle\overline{\Psi(t_{L})}\rangle^{\rm azimuthal}$
in (M$_{\sun}$ Gyr$^{-1}$ pc$^{-2}$).
{\it{Left}}: BSP iteration number 1. {\it{Right}}: BSP iteration number 3.
The $y$-axis indicates the radius ($R$) measured from the centre of the object in kpc.}
~\label{fig6}
\end{figure*}

\begin{figure}
\centering
\includegraphics[width=\columnwidth]{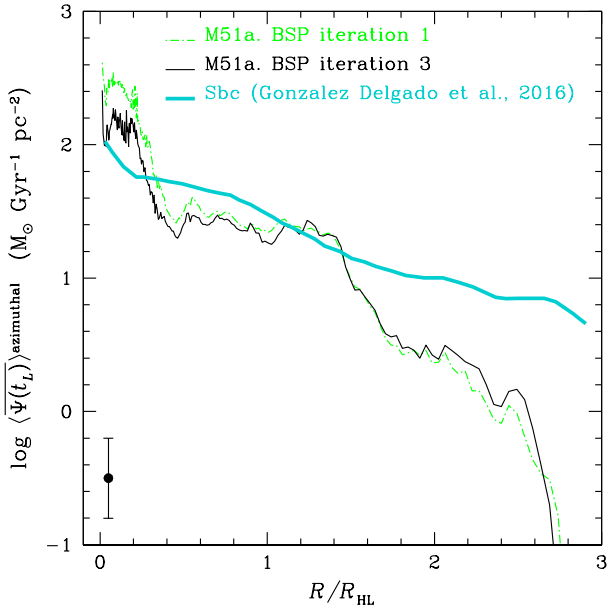}
\caption[fig7.eps]{Azimuthally averaged SFR at $t_{L}=0$, in half-light radius, $R_{\rm HL}$, units.
For M51a, the half-light radius is estimated from the SDSS $g$-band similarly to~\citet{gond14},
resulting in $R_{\rm HL}=4.7$ kpc.
{\it Dashed-dotted green line:} M51a result obtained after BSP iteration number 1.
{\it Solid black line:} M51a result after BSP iteration number 3.
{\it Solid turquoise thick line:}~\citet{gond16} average for Sbc galaxies.
}
~\label{fig7}
\end{figure}

For each value of $t_{L}$ we have obtained a 2D map of the SFR (a total of 137 maps).
In Figure~\ref{fig5} we show the present-day SFR-map for BSP iterations number 1 and 3, respectively.
These SFR-maps were deprojected assuming an inclination angle of $20\degr$, 
and a position angle of $172\degr$~\citep{ler08}.
From these deprojected SFR-maps we then obtained the azimuthally averaged SFR
$\langle\overline{\Psi(t_{L})}\rangle^{\rm azimuthal}$, for annuli at diverse radii ($R$).
The results for all the radial profiles are displayed in Figure~\ref{fig6},
where we present the outcome for BSP iterations 1 and 3, in the left and
right panels, respectively.
Both results show a decreasing SFR with radius. The same behaviour (see Figure~\ref{fig7})
was obtained by~\citet{gond16}
with IFU data from the Calar Alto Legacy Integral Field Area survey~\citep[CALIFA,][]{san12}.
These results have also been corroborated by independent studies~\citep[see e.g.,][]{pil12,nel16}.
In the case of M51a, we also find that the SFR activity is more localised at shorter radii
and look-back times of $1<t_{L}<5$ Gyr, in BSP iteration number 3;
conversely, the maximum-likelihood result (BSP iteration number 1)
entails a wider spread in $t_{L}$.

\section{Discussion}

\subsection{SFH parameterization}~\label{SFH_paramreti}

The recovery of the SFH of individual galaxies is essential to understand
their evolution in the universe~\citep[e.g.,][]{cid05,ocv06,toj07,toj09,wei14,iye17,will17}.
Commonly assumed simple forms of the SFH are a constant SFR ($\Psi$)
with an abrupt cutoff at some time, and an exponentially declining function of time with
a sharp rise at the beginning~\citep[e.g.,][]{tin72,bru83}.
Recent studies have proposed functional forms that include a period of rising
$\Psi$ (not necessarily sharp) at early times, followed by a peak of SF,
and then a decline in $\Psi$ (not necessarily exponential)
until the present-day value~\citep[e.g.,][]{gla13,pac16}.
\citet{gav02} suggested a `delayed-exponential' or `\`a la Sandage' SFH,
with a delayed rise of $\Psi$ up to a maximum, followed by an exponential decrease.
\citet{mar10} propounded an exponentially increasing $\Psi$ to describe the early
SFH of $z\sim 2$ galaxies.
\citet{beh13} posited that the best-fit for various
constraints on individual histories is a double power-law with four free parameters.
Assuming that the SFH of individual galaxies follows a functional form similar to
the cosmic star formation rate density~\citep[symmetric with logarithmic time, e.g.,][]{mad98},
log-normal SFHs have also been suggested~\citep[e.g.,][]{gla13,dre16,die17}.

\begin{figure}
\centering
\includegraphics[width=\columnwidth]{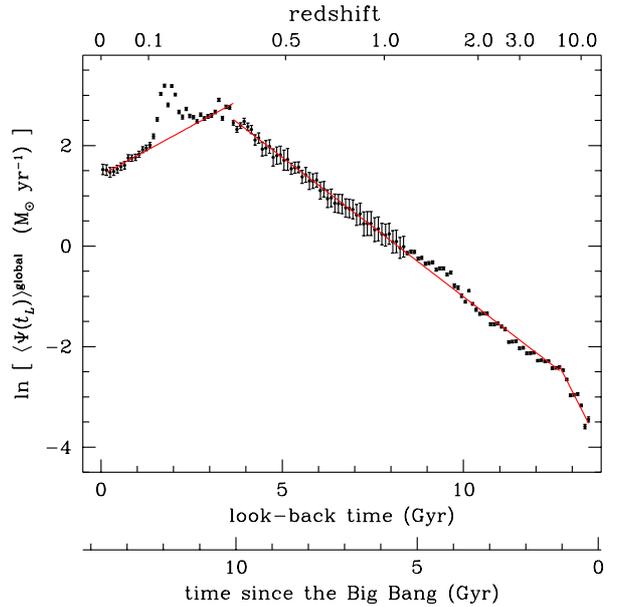}
\caption[fig8.eps]{Fits (solid red lines, see Table~\ref{tbl-1}) to
the SFH of M51a after BSP iteration number 3 (black points).
Error bars represent the absolute error in $\langle\Psi(t_{L})\rangle^{\rm global}$,
propagated for the natural logarithm function ($\ln$).}
~\label{fig8}
\end{figure}

As shown in Figure~\ref{fig4}, for BSP iteration number 1 (maximum-likelihood estimate),
$\langle\Psi(t_{L})\rangle^{\rm global}$ rises from the Big Bang until
$\sim$ 4 Gyr ago, then stays fairly constant
until about 2.5 Gyr ago when it starts to decline.
For BSP iteration number 3 the SFH can be roughly characterised by
three exponential time periods, as illustrated by Figure~\ref{fig8}.
For each of these periods we fit $\Psi(t)$ as

\begin{equation}~\label{exp_SFR}
      \Psi(t) = A \exp\left(-\frac{t-t_{0}}{\tau}\right),
\end{equation}
where $t_{0}$, $t$ are the beginning and ending times
of the periods in time elapsed since the Big Bang,
related to the look-back time by $t_{L} $(Gyr)$= 13.721 - t$,
and $\tau$ is the characteristic $e$-folding time.
The fitted parameters for each time period
are given in Table~\ref{tbl-1}. The first two time periods
are adequately fitted with an exponentially increasing $\Psi(t)$,
i.e., a negative $e$-folding time~\citep[or `inverted-$\tau$' model,][]{mar10},
whereas the last time period corresponds to $\Psi(t)$ declining
exponentially until the present day. Superimposed on top of the most recent two
exponential segments of $\Psi(t)$ we find three bursts of star
formation at $t\approx$ 9.7, 10.4, and 11.9 Gyr after the Big Bang,
or $t_{L}\approx$ 3.6, 3.3, and 1.8 Gyr,
being the latter the most prominent.
The SFH is not symmetric in logarithmic time.

\begin{table}
	\centering
	\caption{Fitted parameters of equation~\ref{exp_SFR}
        for the time periods, $\Delta t$ (time since the Big Bang),
        shown in Figure~\ref{fig8}.}
	\label{tbl-1}
	\begin{tabular}{cccc} 
		\hline
		$\Delta t$ (Gyr) & A (M$_{\sun}$ yr$^{-1}$) & $t_{0}$ (Gyr) & $\tau$ (Gyr)\\
		\hline
		0.3  - 1.1    &  0.03 &  0.3   & -0.703 \\ 
		1.1  - 10.1   &  0.08 &  1.1   & -1.801 \\ 
	       10.1  - 13.7   & 17.13 & 10.1   & ~2.552 \\ 
		\hline
	\end{tabular}
\end{table}

\begin{figure}
\centering
\includegraphics[width=\columnwidth]{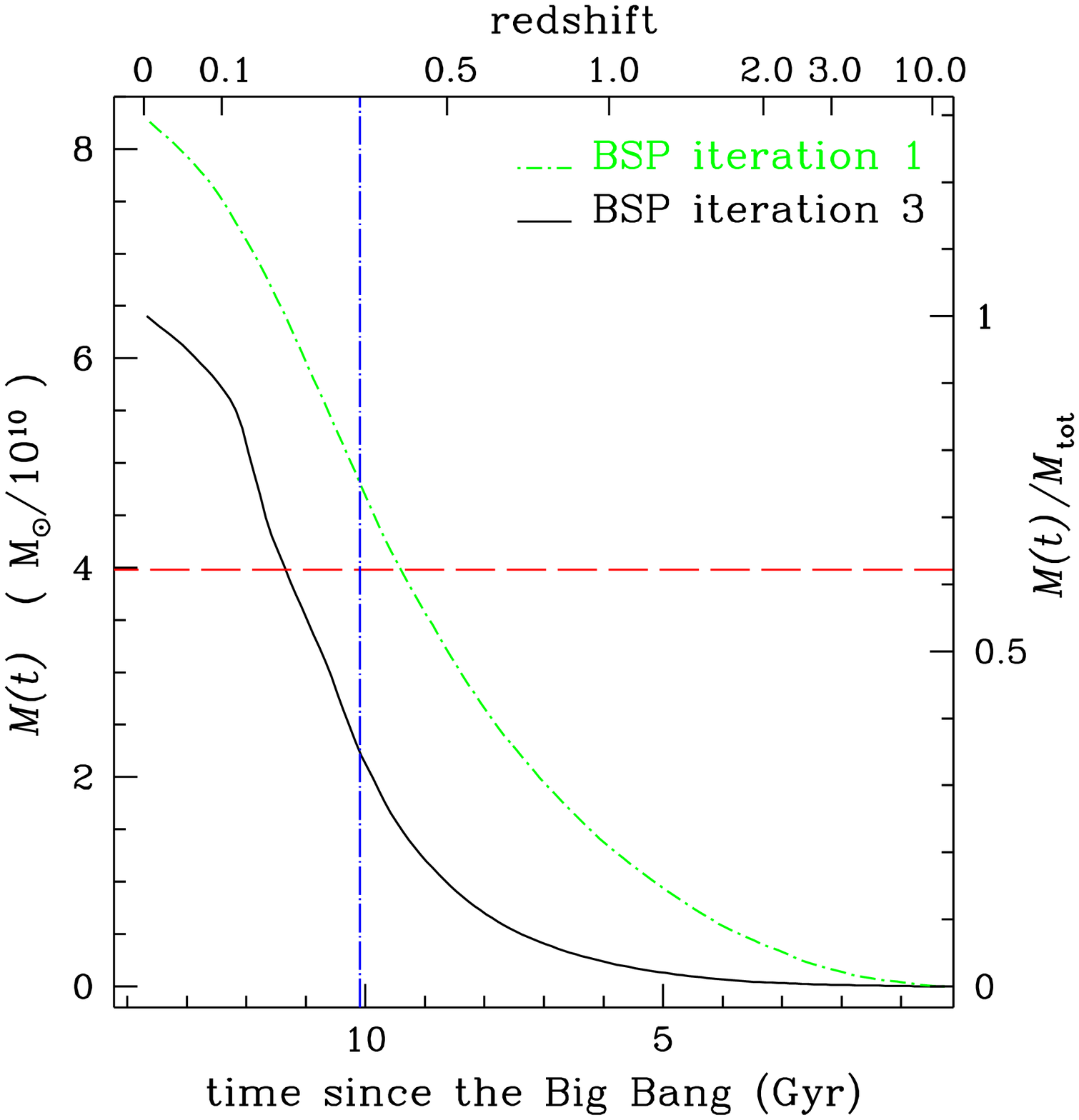}
\caption[fig9.eps]{Mass processed in stars,
$M(t)$ in M$_{\sun}/10^{10}$, for M51a.
{\it Dashed-dotted green line:} result after BSP iteration number 1.
{\it Solid black line:} result after BSP iteration number 3.
{\it Long-dashed (horizontal) red line:} present-day total resolved stellar mass obtained
from the BSP algorithm~\citep{mar17}.
{\it Long-dashed-dotted (vertical) blue line:} $t=10.1$ Gyr.}
~\label{fig9}
\end{figure}

In Figure~\ref{fig9} we show the mass $M(t)$ turned into stars until time $t$, estimated from
\begin{equation}
    M(t) = {\int_{0}^{t} \langle\Psi(t')\rangle^{\rm global} {\rm d}t'}.
\end{equation}

\noindent The resulting $M(t)$ for BSP iteration number 3 has an overall shape
that is qualitatively similar to the one obtained for Milky Way analogues~\citep{die17}.
The present-day resolved stellar mass of M51a
is~$M^{\rm BSP}_{*}=(3.98\pm0.09)\times10^{10}$~M$_{\sun}$,\footnote{
Note that in~\citet{mar17} we adopt a distance of $9.9\pm0.7$~Mpc~\citep{tik09}, resulting
in a total stellar mass of $M^{\rm BSP}_{*}=(5.6\pm0.8)\times10^{10}$~M$_{\sun}$.}
which is marked by the long-dashed (horizontal) red line in Figure~\ref{fig9}.
The present-day value of $M(t)$ is $M_{\rm tot}=6.40\times10^{10}$~M$_{\sun}$, which indicates
that $\sim38$ per cent of this mass has been returned to the ISM,
during the lifetime of the galaxy.
The change from exponentially increasing $\Psi$ (negative $\tau$)
to exponentially decreasing $\Psi$ (positive $\tau$), occurs at $t\sim10.1$ Gyr
after the Big Bang. Interestingly, $\sim35$ per cent of the stellar mass is formed in
the phase of negative $\tau$, and the remaining 65 per cent in the phase
of positive $\tau$ (see Figure~\ref{fig9}).

As a consistency check, we have used~\citet{bru03} to build a galaxy model following
the SFH depicted in Figure~\ref{fig8}. At the present epoch this model reproduces quite
well the integrated luminosity and colours of M51a,
provided that we allow for a dust content as described by the~\citet{cha00}
prescription with $\tau_V \approx 1$. 
Although not a straightforward comparison, since spatially-unresolved results may differ
significantly from resolved ones~\citep[$\sim10-20$ per cent,][]{mar17},
this optical depth is consistent with our BSP
iteration number 3 results,
namely, $\langle\tau_V\rangle=0.93$, and median $\tau_V = 0.78$.
If we ignore the negative $\tau$ segments of $\Psi(t)$ in Figure~\ref{fig8},
we get a model that still resembles
the observed galaxy at the present epoch.
Both models are identical in the UV, but differ in flux in
an amount that increases with spectral wavelength.
The two models also differ in their dust content and age (because of their different onsets of SF).
The model that omits the negative $\tau$, being fainter in the visible and NIR,
will result in a higher estimate of the galaxy mass.

\subsection{The recent SFH of M51a}~\label{recentSFH}

\begin{figure}
\centering
\includegraphics[width=\columnwidth]{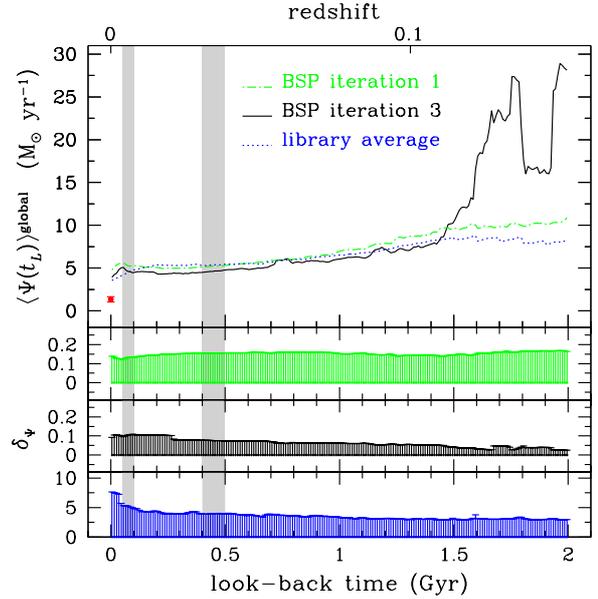}
\caption[fig10.eps]{
The recent 2 Gyr SFH of M51a, obtained with the BSP algorithm
in age bins of 10 Myr. Same nomenclature as in Figure~\ref{fig4}.
Shaded regions in gray indicate time periods of 50-100 and 400-500 Myr ago.
{\it Red point}: the attenuation-corrected $\Psi_{{\rm H}\alpha}$ obtained from
H$\alpha$ and $24\micron$ luminosities~\citep{ken98,ken09}. }
~\label{fig10}
\end{figure}

The SFH of M51a has been the subject of various studies~\citep[e.g.,][]{kal10,kan15}.
\citet{sal00} investigated two numerical models for the M51 interacting system.
One of them predicts a single disc-plane crossing of the companion (M51b);
the other anticipates several crossings.
In both models, the companion's disc-plane crossing responsible for the current spiral structure
occurred 400-500 Myr ago. On the other hand, the multiple crossing model
predicts another more recent disc-plane crossing about 50-100 Myr ago.
From the observational point of view, the
cluster formation rate in M51a also presents a large increase 50-70 Myr ago~\citep{bas05,gie05}.
In Figure~\ref{fig10} we present the most recent 2 Gyr of the SFH obtained
with the BSP algorithm in age bins of 10 Myr (10 times narrower than in Figure~\ref{fig4}). 
We have indicated the time periods of 50-100, and 400-500 Myr ago
with shaded regions.
Our results (BSP iteration number 3) indicate an increase of $\sim29$ per cent in
$\langle\Psi(t_{L})\rangle^{\rm global}$ 50-100 Myr ago, compared to the present-day value,
$\langle\Psi(t_{L})\rangle^{\rm global}=3.9\pm0.4$~M$_{\sun}$ yr$^{-1}$.
However, we find no SF burst 400-500 Myr ago.

Also, in Figure~\ref{fig10} (red point) we show the present-day attenuation-corrected
(for extinction internal to the galaxy itself)
$\Psi_{{\rm H}\alpha}$, derived from the combination of H$\alpha$ and $24\micron$ luminosities.
We adopt the~\citet{ken98} and~\citet[][their equation 12]{ken09} calibrations,
along with the~\citet{ken08} H$\alpha$ + [\ion{N}{ii}], and~\citet{dal07} $24\micron$
integrated fluxes.
We correct for our adopted distance and Galactic reddening,
and for [\ion{N}{ii}] emission~\citep{ken08}. We divide the resulting $\Psi_{{\rm H}\alpha}$
by $1.7\pm0.3$ to convert to a~\citet{cha03} IMF, since~\citet{ken98} calibrations
correspond to a~\citet{sal55} IMF.
We estimate a value of $\Psi_{{\rm H}\alpha}=1.34\pm0.32$ $M_{\sun}$ yr$^{-1}$.
Our previously derived $\langle\Psi(t_{L})\rangle^{\rm global}$ is $\sim3$ times
bigger than $\Psi_{{\rm H}\alpha}$. The measurements are different by $3.55\sigma$.\footnote{
For the SSAG-2015 (see appendix~\ref{SFH_prior}) we obtain
$\langle\Psi(t_{L})\rangle^{\rm global}=3.1\pm0.8$~M$_{\sun}$ yr$^{-1}$. In this
case the measurements differ only by $1.57\sigma$.}

\section{A different choice of prior PDF}

\begin{figure} 
\centering
\includegraphics[width=\columnwidth]{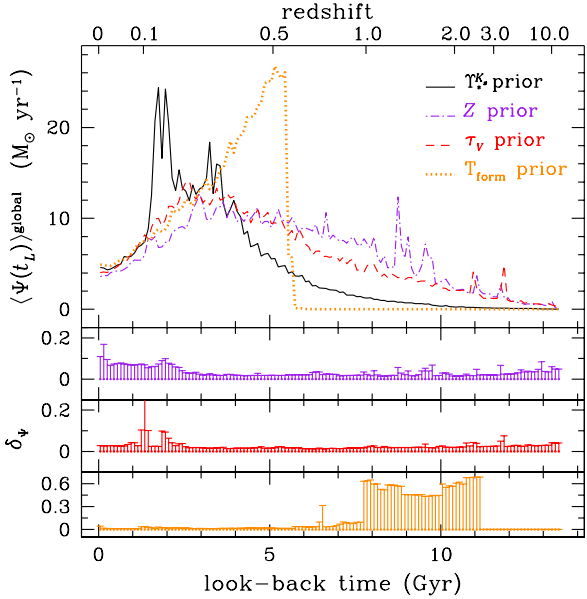}
\caption[fig11.eps]{
Top panel: same as Figure~\ref{fig4} for different prior PDFs.
All SFH curves correspond to results after BSP iteration number 3.
{\it Solid black}, {\it dashed-dotted purple}, {\it dashed red},
and {\it dotted orange} lines: mass-to-light ratio ($\Upsilon^{K_{s}}_{*}$),
stellar metallicity ($Z$), dust ($\tau_V$), and stellar age (T$_{\rm form}$)
priors, respectively.
Second, third, and fourth from top panels: relative errors,
$\delta_{\Psi}=\frac{\sigma_{\Psi}}{\psi}$, for the $Z$,
$\tau_V$, and T$_{\rm form}$ priors, respectively.}
~\label{fig11}
\end{figure}

\begin{figure} 
\centering
\includegraphics[width=\columnwidth]{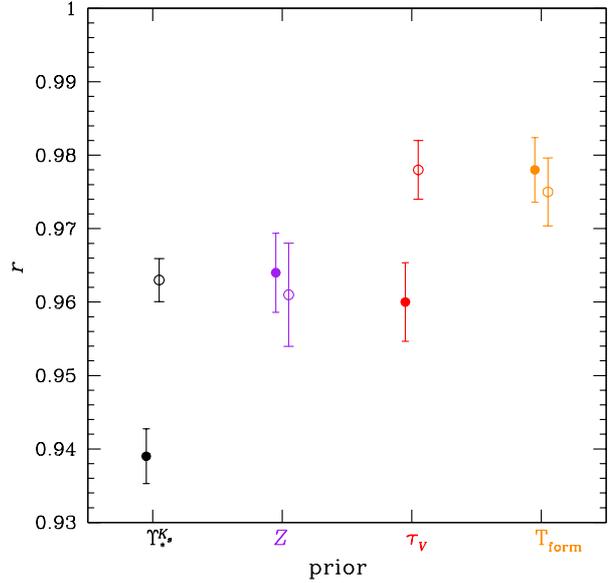}
\caption[fig12.eps]{
{\it Solid symbols}: Pearson correlation coefficient, $r$ (see equation~\ref{pcc}),
between the mass-maps resulting from BSP iterations number 1 and 3.
The adopted prior PDF is indicated on the $x$-axis
for the mass-to-light ratio $\Upsilon^{K_{s}}_{*}$,
the metallicity $Z$,
the dust content $\tau_V$,
and the stellar age T$_{\rm form}$, respectively.
{\it Open symbols}: a radially varying prior PDF is assumed
for BSP iteration number 2 (see section~\ref{radial_prior}).}
~\label{fig12}
\end{figure}

Up to this point we have applied the~\citet{mar17} BSP algorithm
to infer the SFH of M51a. This is done by assuming a prior PDF
for $\Upsilon^{K_{s}}_{*}$ in equation~\ref{eq_PUpsilon}. In this section
we explore the use of prior PDFs for other parameters, for instance,
the stellar metallicity, $Z$, the dust content characterised by $\tau_V$,
and the stellar age characterised by T$_{\rm form}$.\footnote{
Throughout the manuscript `BSP' refers to a prior PDF
for $\Upsilon^{K_{s}}_{*}$, unless otherwise indicated.}
For this purpose, equation~\ref{eq_PUpsilon} takes the form

\begin{equation}~\label{eq_PX}
P(\mathbf{X} \mid C) \propto \exp \left(-\frac{\chi^{2}}{2}\right)
\exp \left(-\frac{1}{2}\left[\frac{\mathbf{X}^{\rm prior} - \mathbf{X}} 
{\sigma_{\mathbf{X}}} \right]^{2}\right),
\end{equation}\\

\noindent where $\mathbf{X}$ is replaced by $Z$, $\tau_V$, or T$_{\rm form}$, and
$\sigma_{\mathbf{X}}\approx\sigma_{\rm mag}\mathbf{X}^{\rm prior}$.
We apply BSP in a similar manner to that used with the
$\Upsilon^{K_{s}}_{*}$ prior PDF. The only difference is when
moving from iteration number 2 to iteration number 3: we compute
the required interpolated structure directly from the
`backbone' $\mathbf{X}$ pixels, instead of interpolating it
from the `backbone' mass pixels~\citep[cf.][]{mar17}.
The results of this exercise are shown in Figure~\ref{fig11}.
The SFH curves for $Z$ and $\tau_V$ follow an almost identical
behaviour to the one obtained from the maximum-likelihood
estimate (BSP iteration number 1). Contrariwise, the T$_{\rm form}$
SFH curve presents a shape very similar to our SFH templates (see Figure~\ref{fig3}),
with T$_{\rm form}\sim5.5$ Gyr. This result is not surprising, since
the median T$_{\rm form}$ value of the whole disk, after BSP iteration
number 1, is T$_{\rm form}=5.71$ Gyr. In this manner, we are basically
recovering the prior SFH we assumed before.
A difference between the results obtained with the $\Upsilon^{K_{s}}_{*}$
and the T$_{\rm form}$ priors, respectively, is that the latter still produces mass-maps
with a filamentary structure. We have corroborated this effect qualitatively
by visual inspection, and quantitatively by calculating the normalised
Pearson correlation coefficient between two mass-maps,

\begin{equation}~\label{pcc}
     r =   \frac{\sum\limits_{j}\sum\limits_{i}(F_{ij}-\bar{F})(G_{ij}-\bar{G})}
          {\sqrt{\sum\limits_{j}\sum\limits_{i}(F_{ij}-\bar{F})^2}
           \sqrt{\sum\limits_{j}\sum\limits_{i}(G_{ij}-\bar{G})^2}},
\end{equation}

\noindent where $F_{ij}$ is the stellar mass surface density, $\rho_{*}$,
of the $i^{\rm th},j^{\rm th}$ pixel in the first mass-map,
$G_{ij}$ is the $\rho_{*}$ of the $i^{\rm th},j^{\rm th}$ pixel in the second mass-map,
$\bar{F}$ is the mean $\rho_{*}$ of the first mass-map,
and $\bar{G}$ is the mean $\rho_{*}$ of the second mass-map.
We estimate $r$ between the resulting mass-maps for BSP iterations number 1
and 3, adopting the $\Upsilon^{K_{s}}_{*}$,
$Z$, $\tau_V$, and T$_{\rm form}$ priors. The uncertainties in $r$ are
estimated via bootstrap methods~\citep[e.g.,][]{bha90}.
The results are plotted in Figure~\ref{fig12}, where a value of $r=1$ would indicate a perfect
match between the spatial structure of both mass-maps. As expected,
the $r$ value for the $\Upsilon^{K_{s}}_{*}$ prior PDF indicates
a greater discrepancy between the mass-maps when compared to the $r$ values
obtained when adopting the $Z$, $\tau_V$, or T$_{\rm form}$ prior PDFs.
As stressed out by~\citet{mar17}, an advantage of using a spatial structure prior,
by means of a $\Upsilon^{\rm NIR}_{*}$ prior, with our BSP algorithm,
is its independence from SPS model parameters, e.g., SFH, metallicity, dust, age, etc.

\subsection{A radially varying prior PDF}~\label{radial_prior}

\begin{figure} 
\centering
\includegraphics[width=\columnwidth,angle=-90,scale=0.8]{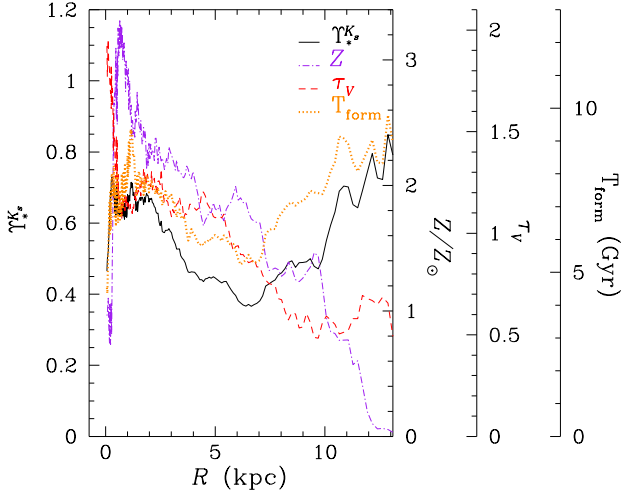}
\caption[fig13.eps]{
Azimuthally averaged $\mathbf{X}$ parameters vs. radius, $R$ (kpc),
for M51a deprojected maps, after BSP iteration number 1.
{\it Solid black line:} mass-to-light ratio ($\Upsilon^{K_{s}}_{*}$).
{\it Dashed-dotted purple line:} stellar metallicity $Z/Z_{\sun}$.
{\it Dashed red line:} dust ($\tau_V$).
{\it Dotted orange line:} stellar age (T$_{\rm form}$). }
~\label{fig13}
\end{figure}

\begin{figure} 
\centering
\includegraphics[width=\columnwidth]{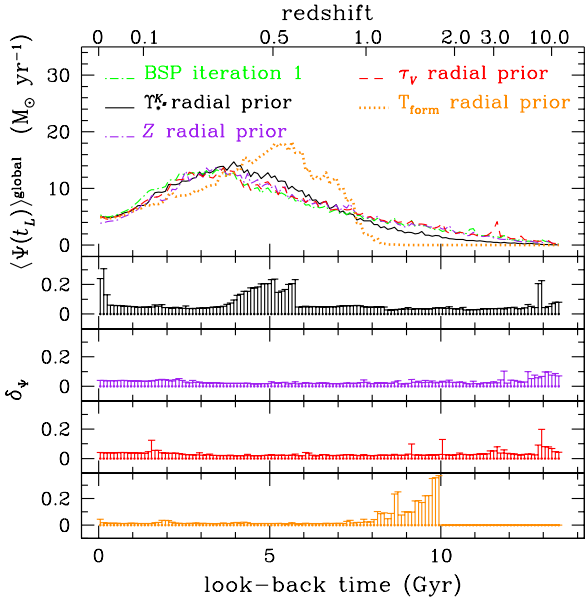}
\caption[fig14.eps]{
Resulting SFHs of M51a after a radially varying prior PDF (see section~\ref{radial_prior}).
{\it Dashed-dotted green line:} result after BSP iteration number 1.
{\it Solid black}, {\it dashed-dotted purple}, {\it dashed red}, and
{\it dotted orange} lines indicate the SFH after BSP iteration number 3,
assuming a radially varying $\Upsilon^{K_{s}}_{*}$,
$Z$, $\tau_V$, and T$_{\rm form}$ prior in BSP iteration number 2, respectively.
Second, third, fourth, and fifth from top panels: relative errors for the
$\Upsilon^{K_{s}}_{*}$, $Z$, $\tau_V$, and T$_{\rm form}$ SFHs, respectively.}
~\label{fig14}
\end{figure}

As described in section~\ref{analysis}, the second BSP iteration assumes a constant
stellar mass-to-light ratio $\Upsilon_{*}$, in the NIR, for the entire disk,
i.e., a constant $\Upsilon_{*}^{\rm prior}$, or a constant $\mathbf{X}$ parameter,
in equations~\ref{eq_PUpsilon} or~\ref{eq_PX}, respectively. 
In this section we explore the use of a radially varying $\mathbf{X}$.
For this purpose we use the $\mathbf{X}$ radial profiles resulting
from BSP iteration number 1. However, we must remember that the mass-map generated
from this maximum-likelihood fit has a bias in its spatial structure. The radial profiles
for $\Upsilon^{K_{s}}_{*}$, $Z$, $\tau_V$, and T$_{\rm form}$ are shown in Figure~\ref{fig13}.
We use these radial profiles to generate a surface of revolution
for each $\mathbf{X}$ parameter. We synthesise some images from these
surfaces, project them to mimic the disk orientation,
and use them in BSP iteration number 2,
instead of the constant plane previously assumed.
The resulting SFHs of this test are shown in Figure~\ref{fig14}.
We find a tendency to follow the curve for BSP iteration number 1
in all the parameters, which is more accentuated for the $Z$ and $\tau_V$ curves.
The Pearson correlation coefficient, $r$ (see Figure~\ref{fig12}, open symbols),
between the mass-maps of BSP iterations number 1 and 3 indicate that these assumptions
do not improve the mass-maps more than assuming a constant $\Upsilon^{K_{s}}_{*}$
for BSP iteration number 2.

\section{Conclusions}

We have introduced a novel technique to determine the SFH of a galaxy.
The method is based on spectral fitting pixel by pixel the resolved image
of the galaxy in various photometric bands (UV to NIR), using our
BSP algorithm~\citep{mar17}. We obtain the SFH for each pixel in the galaxy,
i.e., the SFH-map of the galaxy, an image of the galaxy resolved in space and time.
We can thus characterise the underlying shape of the SFH  from the Big Bang to
the present day, together with individual episodes, or bursts, of star formation.
We have applied this technique to M51a and find that its global SFH consists of
an exponentially increasing SFR $\Psi(t)$ lasting until $\approx$ 10 Gyr
after the Big Bang, followed by an exponentially decreasing 
$\Psi(t)$ until the present day, with a main burst of star formation
superimposed during the declining phase.
These results show that, while the SFH of each individual pixel of a
galaxy can be adequately fitted by an exponentially decaying $\Psi(t)$,
the global SFH of the galaxy may behave differently in time.
Nevertheless, we recognise that as in every problem solved with Bayesian statistics,
the solution will reflect the physical properties of the
prior used (SSAG), and we do not rule out that a radical change in the SSAG properties,
may result in a drastic change for the SFH of M51a.
The SFH of a disk galaxy has to be compatible with the present-day stellar
mass distribution as inferred from NIR images. This is only
achieved when a mass-to-light ratio prior PDF is adopted in BSP.
When applied to a larger sample of galaxies this method can help us constrain current
models of galaxy evolution, as well as the cosmic SFH~\citep[e.g.,][]{hea04,mad14}.

\section*{Acknowledgements}

We acknowledge the reviewer for important comments and suggestions.
EMG acknowledges the remote use of the computer `galaxias' at IRyA, UNAM.
GB acknowledges support for this work from UNAM through grant PAPIIT IG100115.
RAGL thanks DGAPA, UNAM, for support through the PASPA program.
We appreciate the usefulness of the {\it GALEX} Atlas of Nearby Galaxies
website, \url{https://archive.stsci.edu/prepds/galex_atlas/.}
The SDSS-III web site is \url{http://www.sdss3.org/}.








\appendix

\section{The implicit SFH\lowercase{s} prior}~\label{SFH_prior}

\begin{figure} 
\centering
\includegraphics[width=\columnwidth]{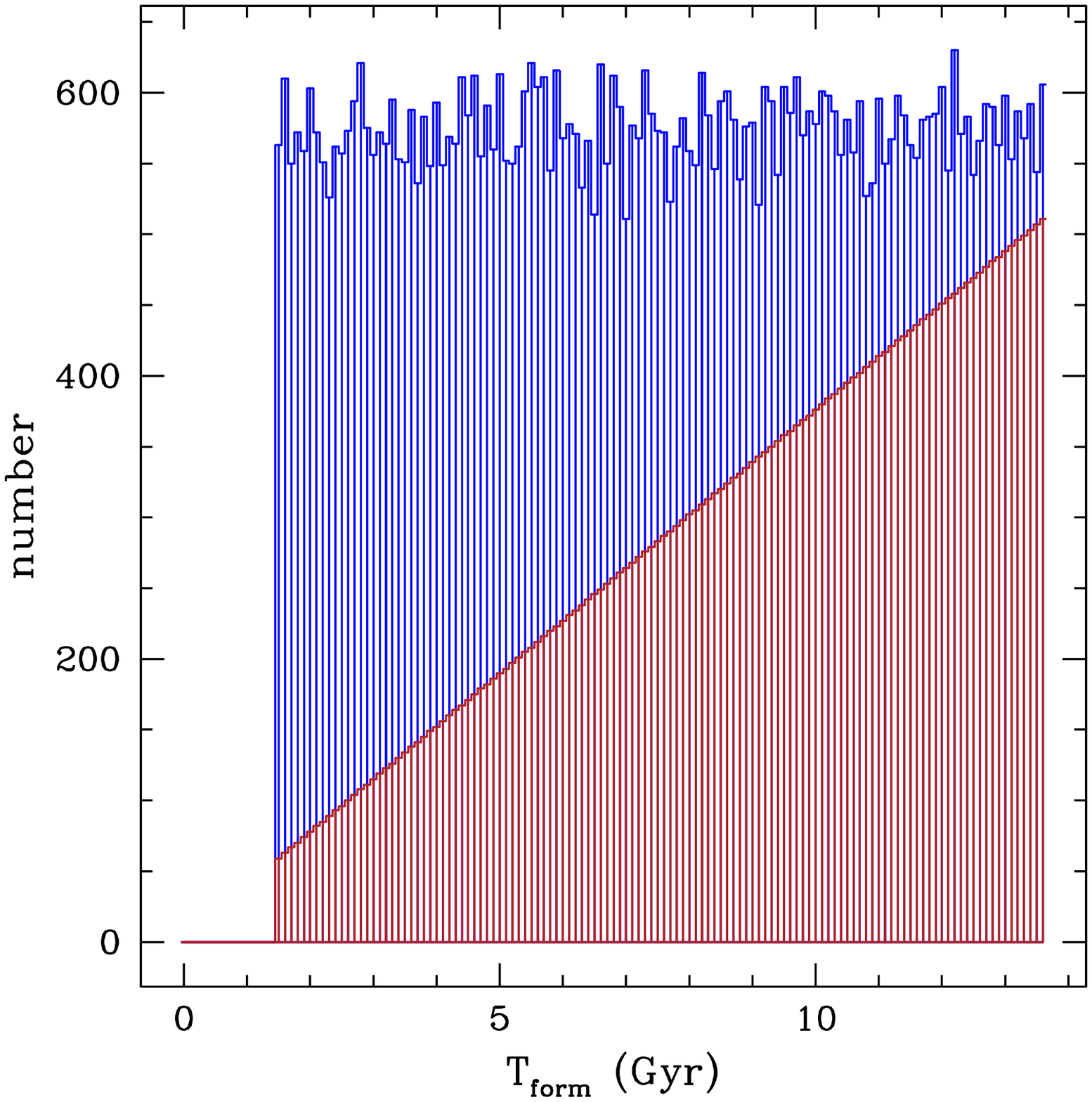}
\caption[figA1.eps]{
T$_{\rm form}$ histograms for the full SSAG-2017 library (blue)
and the subset-I library (dark red).}
~\label{figA1}
\end{figure}

In this appendix we discuss the implicit prior placed on the SFHs by the
adopted SSAG-2017 library. The average of all SFHs,
$\langle\Psi(t_{L})\rangle^{\rm library~average}$ (dotted blue line in Figure~\ref{fig4}),
is dominated by the PDF of the T$_{\rm form}$ parameter, which has a roughly uniform
distribution between 1.5 and 13.7 Gyr (see Figure~\ref{figA1}, blue line histogram).
The T$_{\rm form}$ PDF is correlated to the form of the
$\langle\Psi(t_{L})\rangle^{\rm library~average}$ curve in the following manner.
At look-back times $t_{L}\sim$13.7 Gyr
only a few templates contribute to $\langle\Psi(t_{L})\rangle^{\rm library~average}$. At look-back times
$t_{L}\lesssim$13.7 Gyr the $\langle\Psi(t_{L})\rangle^{\rm library~average}$ has contributions from
the accumulated $\langle\Psi(t_{L})\rangle$ for $t_{L}\sim$13.7 Gyr,
and the corresponding $\langle\Psi(t_{L})\rangle$ for $t_{L}\lesssim$13.7 Gyr. 
This has the consequence of increasing $\langle\Psi(t_{L})\rangle^{\rm library~average}$
as $t_{L}$ decreases within the interval $1.5 \lesssim t_{L} \lesssim 13.7$ Gyr. At $t_{L}\lesssim1.5$ Gyr,
$\langle\Psi(t_{L})\rangle^{\rm library~average}$ decreases together with $t_{L}$, as a consequence of
the T$_{\rm form}$ PDF.

\begin{figure} 
\centering
\includegraphics[width=\columnwidth]{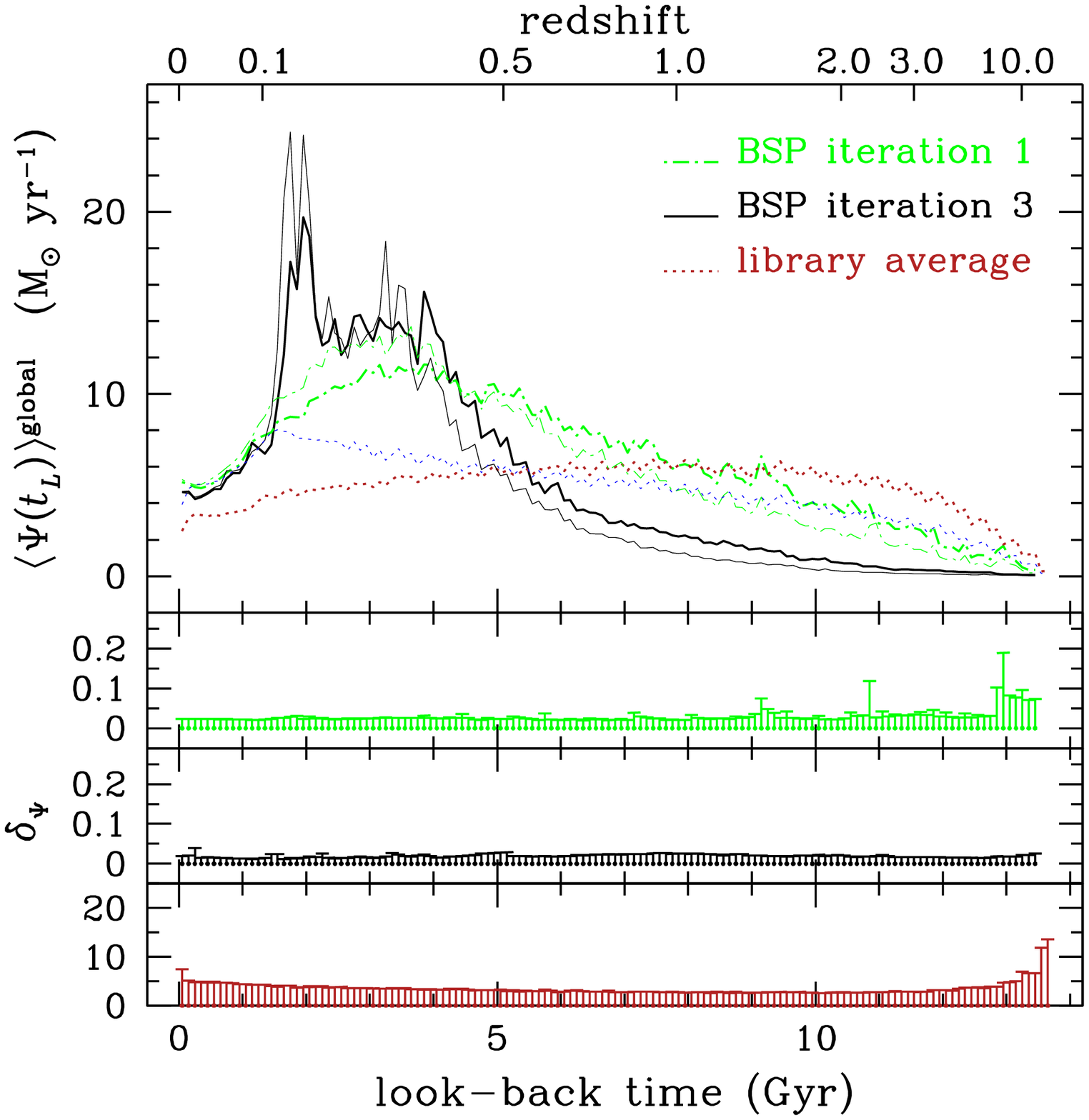}
\caption[figA2.eps]{
Same as Figure~\ref{fig4} for the T$_{\rm form}$ PDF
shown by the dark red histogram in Figure~\ref{figA1}.
{\it Dotted dark red line:} the average of all SFHs in the subset-I library,
multiplied by the present-day resolved stellar mass.
For comparison purposes,
the thinner lines show the same curves plotted in Figure~\ref{fig4}.}
~\label{figA2}
\end{figure}

We select from the SSAG-2017 library the subset shown by the dark red histogram in Figure~\ref{figA1}.
This T$_{\rm form}$ PDF was chosen because it is radically different from the original
SSAG-2017 library (blue histogram).
The resulting library consists of $\sim3.5\times10^4$ templates.
With this subset-I library, we apply BSP in an identical manner as before.
The results are shown in Figure~\ref{figA2}. In the same figure we
also plot our SSAG-2017 (Figure~\ref{fig4}) results with thinner lines.
In order to verify the equality (or not) between both results we
perform a Kolmogorov-Smirnov (KS) test.
The KS test compares the cumulative empirical distribution function
of some sample data with the expected distribution (or expected SFH curve in our case).
The KS statistic, $D$, is used to compute the probability $p_{\rm value}$,
which gauges the evidence against dissimilar distributions.
Typically, when $p_{\rm value}>0.05$, there is not enough evidence to conclude
that the two distributions differ from each other,
whereas $p_{\rm value}<0.05$ indicates distinct distributions.
In Table~\ref{tbl-A1} we show the results of the KS test between
the full SSAG-2017 and the subset-I libraries. 
For the BSP curves, the $p_{\rm value}$ probabilities indicate analogous distributions,
while for the library average curves, the $p_{\rm value}$ point to different distributions,
thus demonstrating no correlation between our implicit SFHs prior and BSP results.

\begin{table}
	\centering
	\caption{KS test results between SSAG-2017 and other libraries.}
	\label{tbl-A1}
	\begin{tabular}{lccc} 
		\hline
		Library      & SFH curve        & $p_{\rm value}$ & Figure\\
		\hline
		subset-I     & BSP-1            & 0.1379          & \ref{figA2}\\
		             & BSP-3            & 0.1379          & \\
		             & library average  & 0.0043          & \\
		\hline
		subset-II    & BSP-1            & 0.0763          & \ref{figA3}\\
		(no bursts)  & BSP-3            & 0.0469          & \\
		             & library average  & 0.0018          & \\
		\hline
		subset-III   & BSP-1            & 0.0762          & \ref{figA4}\\
		(bursts only)& BSP-3            & 0.3752          & \\
		             & library average  & 0.0001          & \\
		\hline
		SSAG-2015    & BSP-1            & 0.0010          & \ref{figA5}\\
		             & BSP-3            & 0.2348          & \\
		             & library average  & 0.5586          & \\
		\hline
		subset-IV    & BSP-1            & 0.0281          & \ref{figA6}\\
		             & BSP-3            & 0.3752          & \\
		             & library average  & 1.0000          & \\
		\hline
	\end{tabular}
\end{table}

\begin{figure} 
\centering
\includegraphics[width=\columnwidth]{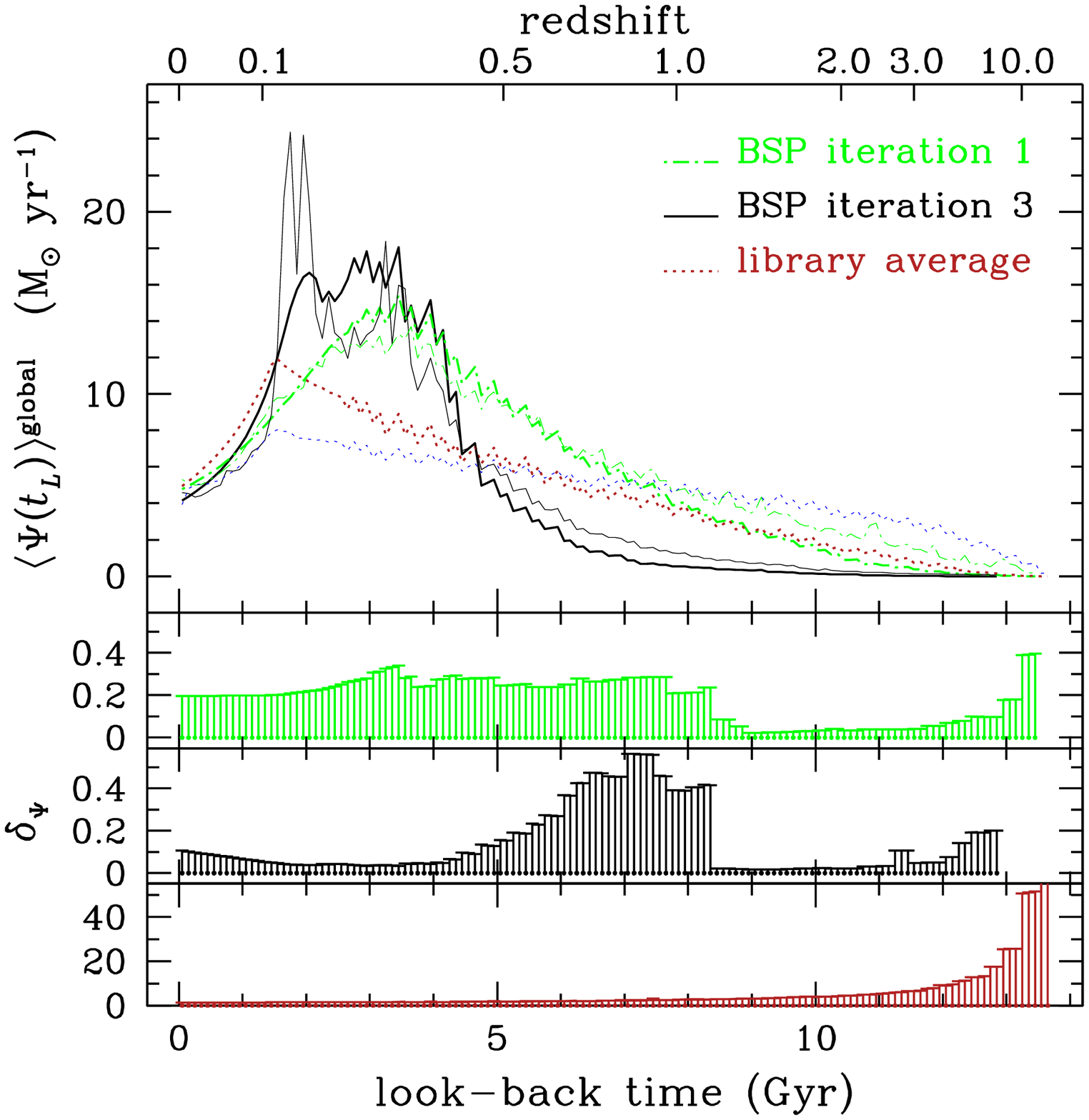}
\caption[figA3.eps]{
Same as Figure~\ref{fig4} for the subset-II (no bursts) library.
{\it Dotted dark red line:} the average of all SFHs in the subset-II library,
multiplied by the present-day resolved stellar mass.
For comparison purposes,
the thinner lines show the same curves plotted in Figure~\ref{fig4}.}
~\label{figA3}
\end{figure}

\begin{figure} 
\centering
\includegraphics[width=\columnwidth]{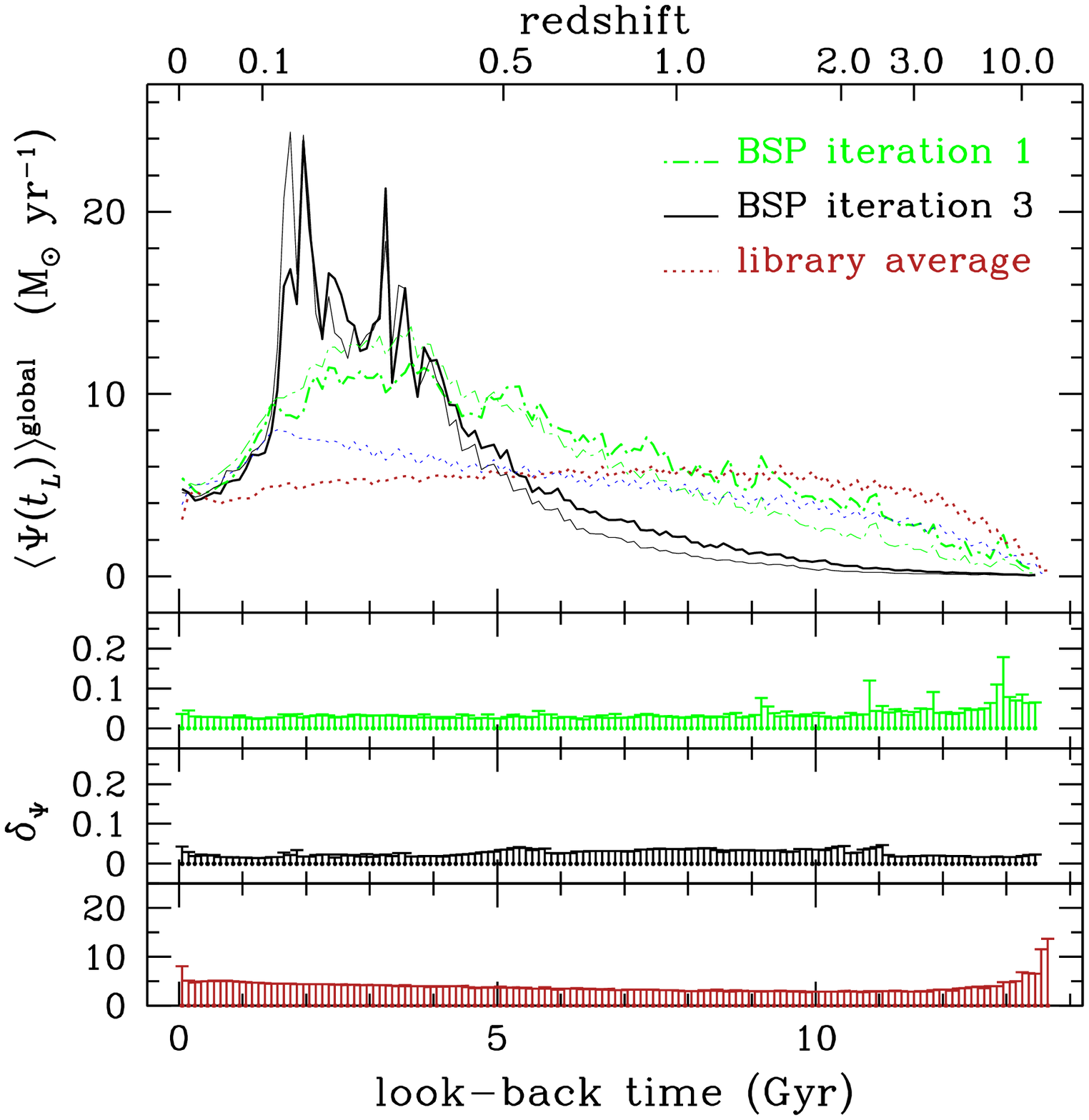}
\caption[figA4.eps]{
Same as Figure~\ref{fig4} for the subset-III (bursts only) library.
{\it Dotted dark red line:} the average of all SFHs in the subset-III library,
multiplied by the present-day resolved stellar mass.
The thinner lines show the same curves plotted in Figure~\ref{fig4}.}
~\label{figA4}
\end{figure}

\begin{figure} 
\centering
\includegraphics[width=\columnwidth]{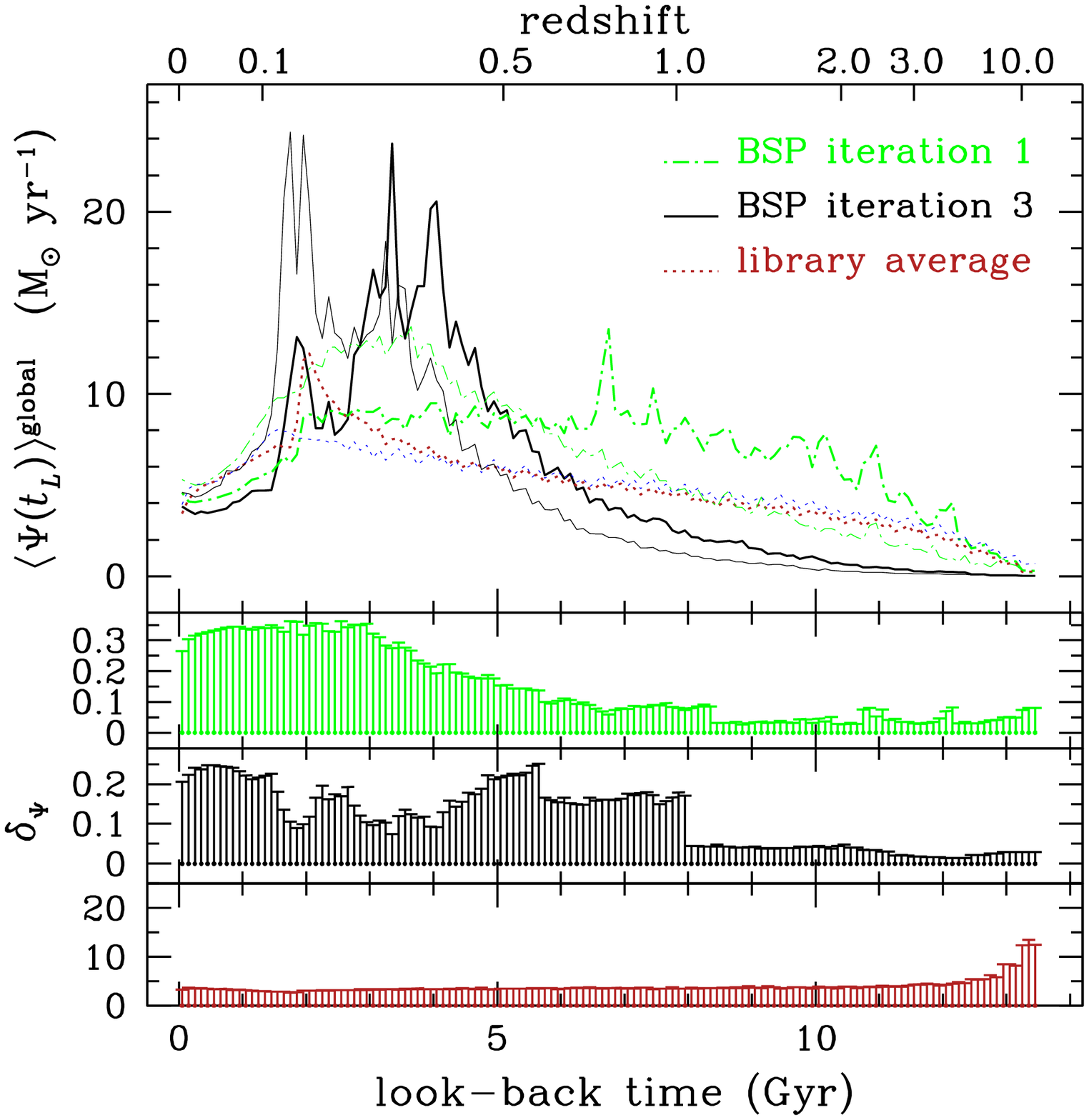}
\caption[figA5.eps]{
Same as Figure~\ref{fig4} for the SSAG-2015 library.
{\it Dotted dark red line:} the average of all SFHs in the SSAG-2015 library,
multiplied by the present-day resolved stellar mass.
The thinnest lines show the same curves plotted in Figure~\ref{fig4}.}
~\label{figA5}
\end{figure}

\begin{figure} 
\centering
\includegraphics[width=\columnwidth]{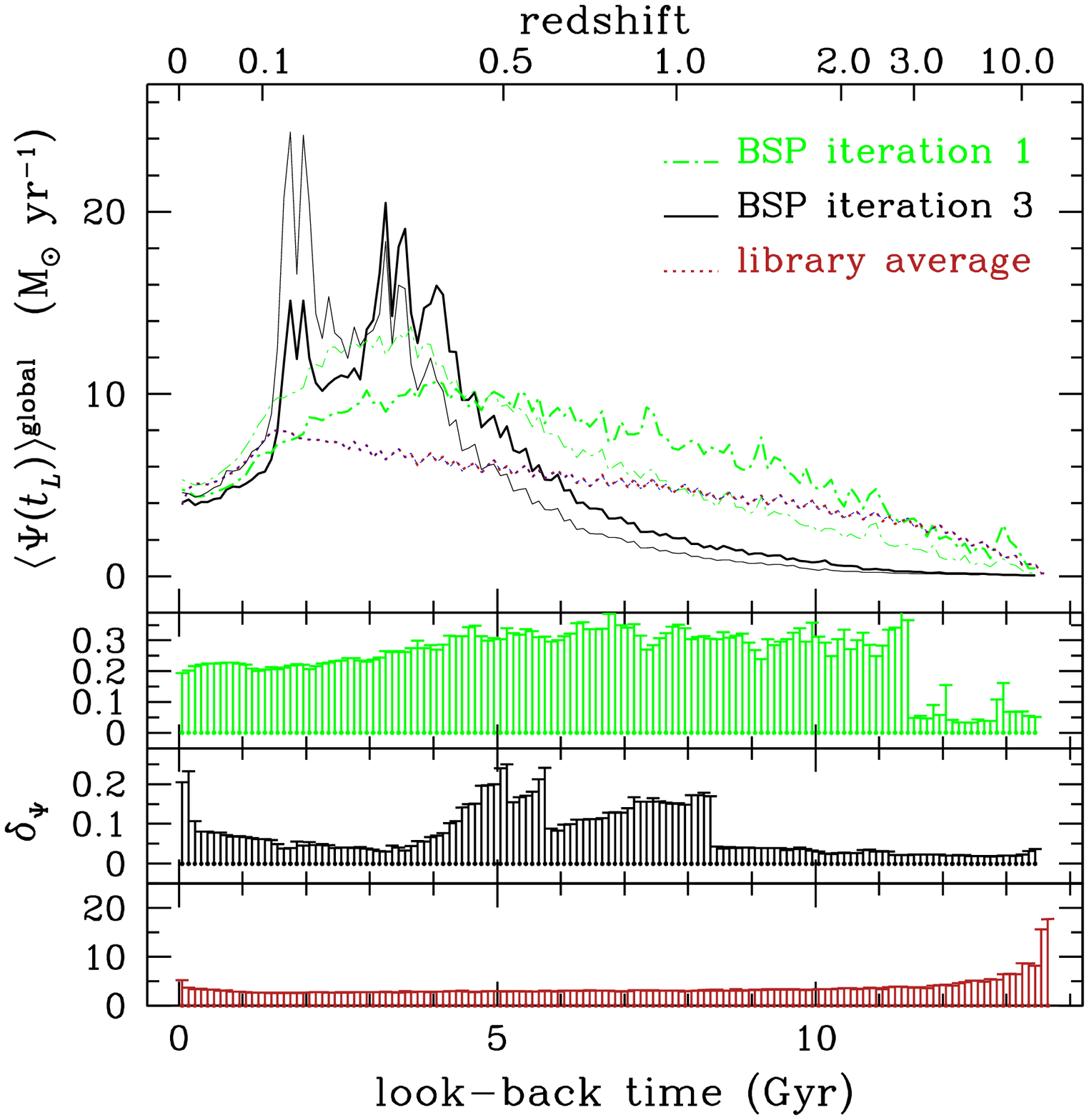}
\caption[figA6.eps]{
Same as Figure~\ref{fig4} for the subset-IV library.
{\it Dotted dark red line:} the average of all SFHs in the subset-IV library,
multiplied by the present-day resolved stellar mass.
The thinnest lines show the same curves plotted in Figure~\ref{fig4}.}
~\label{figA6}
\end{figure}

\begin{figure} 
\centering
\includegraphics[width=\columnwidth]{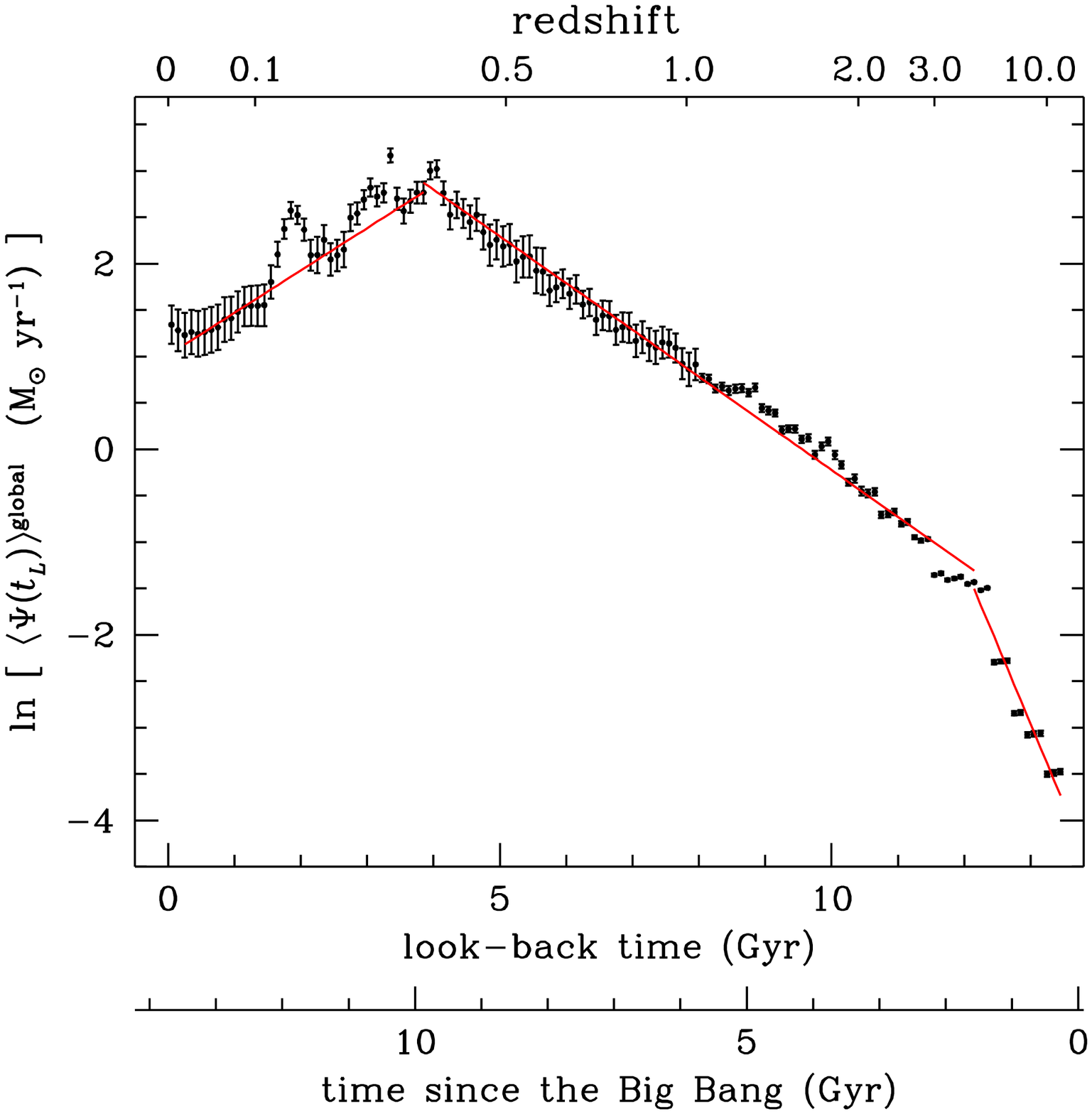}
\caption[figA7.eps]{
Same as Figure~\ref{fig8} for the SSAG-2015 library
(see also Table~\ref{tbl-A2}.)}
~\label{figA7}
\end{figure}

\begin{table}
	\centering
	\caption{Fitted parameters of equation~\ref{exp_SFR}
        for the time periods, $\Delta t$ (time since the Big Bang),
        shown in Figure~\ref{figA7}.}
	\label{tbl-A2}
	\begin{tabular}{cccc} 
		\hline
		$\Delta t$ (Gyr) & A (M$_{\sun}$ yr$^{-1}$) & $t_{0}$ (Gyr) & $\tau$ (Gyr)\\
		\hline
		0.3  - 1.6    &  0.02 & 0.3    & -0.583 \\ 
		1.6  - 9.9    &  0.27 & 1.6    & -1.984 \\ 
		9.9  - 13.5   & 16.00 & 9.9    & ~2.190 \\ 
		\hline
	\end{tabular}
\end{table}

We also compare the results obtained with other three different libraries.
The first one consists in a subset (II) of the SSAG-2017 library,
of all the templates with no superimposed burst of SF.
The second one (subset-III) includes only the SSAG-2017 templates with a superimposed burst of SF.
Additionally, we use the 2015 version of the SSAG as another comparison case.
The SFH curves are shown in Figures~\ref{figA3},~\ref{figA4}, and~\ref{figA5},
for the subset-II, subset-III, and SSAG-2015 libraries, respectively.
In Table~\ref{tbl-A1} we show the KS test results for each case.
As expected, the no-bursts library (subset II) reproduces the general shape
of the SFH curve without the bursts peaks. For the only-bursts library (subset III), the SFH curves
are nearly identical to the SSAG-2017 library case.
For the SSAG-2015 results, the BSP iteration number 1 curve differs from the SSAG-2017 result,
while the BSP iteration number 3 curves reveal similar features but different
$\langle\Psi(t_{L})\rangle^{\rm global}$ amplitudes for both libraries.
This behaviour is mainly due to the different metallicity
ranges, since for SSAG-2015, $Z/Z_{\sun}$ is distributed almost uniformly between 0.02 and 2.5.
To corroborate this effect we separate another subset (IV) of the
SSAG-2017, consisting of all the templates where $0.02 \leqq Z/Z_{\sun} \leqq 2.5$.
The results are shown in Figure~\ref{figA6},
where we can appreciate a very similar outcome as in Figure~\ref{figA5} (see also Table~\ref{tbl-A1}).
By comparing the SSAG-2015 and the subset-IV SFH curves, for BSP iteration number 3,
we obtain $p_{\rm value}=0.9251$. In this manner, a narrower metallicity range can
affect the amplitudes of the SFH curves for certain segments,
but the qualitative behaviour would remain the same.

Finally, in Figure~\ref{figA7} (solid red lines) and Table~\ref{tbl-A2},
we show the fits to the SFH of M51a after BSP iteration number 3,
adopting the SSAG-2015 library. For this case, the turnover
from exponentially increasing $\Psi$ (negative $\tau$)
to exponentially decreasing $\Psi$ (positive $\tau$) occurs at $t\sim9.9$ Gyr.
A very similar value is obtained from the fits shown in Figure~\ref{fig8}, where $t\sim10.1$ Gyr.
In conclusion, the implicit prior PDF of the SFH,
determined by the adopted SPS library, has no relevant effects on the results.


\bsp	
\label{lastpage}
\end{document}